\documentstyle[12pt]{article}
\input epsf

\begin{document}
\begin{titlepage}
\thispagestyle{empty}



\title{ 
\vspace*{-1cm}
\begin{flushright}
{\small CPHT S758.0100}
\end{flushright}
\vspace{2.0cm}
When Is It Possible to Use Perturbation Technique in Field Theory ? }
\vspace{4.0cm}
\author{Tran N. Truong \\
\small \em Centre de Physique Th{\'e}orique, 
{\footnote {unit{\'e} propre 014 du
CNRS}}\\ 
\small \em Ecole Polytechnique \\
\small \em F91128 Palaiseau, France}

\date{May 2000 }

\maketitle

\begin{abstract}
 The vector pion form
factor is used as an example to analyze this question. Given the experimental
radius of the pion, the crucial question is whether perturbative methods could
be used for the effective chiral lagrangian to calculate the pion form factor.
Our analysis shows that the pion rms radius is far too
large (or the related $\rho$ resonant mass is too low) for the perturbation
theory to be valid.

\end{abstract}
\end{titlepage}

\newpage
\section{Introduction}

Given an interaction Lagrangian characterized by a dimensionless coupling constant $g$,
one can usually guess whether perturbation theory can be used by comparing $g^2/4\pi$
with unity. If it is much less than unity, there is a good chance that the perturbation
technique could be used. Effective Chiral Lagrangian gives on the other hand a power
series expansion in momenta and hence the above rule cannot be applied.  It is important
to set up a set of rules to test whether perturbation theory could be used.

 Chiral Perturbation Theory (ChPT) \cite{holstein,Weinberg, GL1, GL2} is a
well-defined perturbative procedure for the Effective Chiral Lagrangian 
 allowing one to calculate systematically low energy phenomenon involving
soft pions. It is now widely used to analyze  
low energy pion physics even in the presence
of  resonance as long as the energy region of interest is sufficiently far from the
resonance. In this scheme, the
unitarity relation is satisfied perturbatively order by order.

 The standard
procedure of testing  ChPT calculation of the pion form factor \cite{Gasser3}, 
which claims
to support the perturbative scheme, is shown here to be unsatisfactory. 
This is so because 
the calculable terms are extremely small, less than 1.5\% of the 
uncalculable terms at an
energy of 0.5 GeV or lower whereas the experimental errors are of the
order 10-15\%. This obvious fact escapes the attention of many  due to  the complexity
of the calculations. 

 We  show how this situation can be dealt with without asking
for a new measurement of the pion form factor with a precision much
better than 1.5\%. This can be done using dispersion relation which establishes 
in a model independent way a relation between the real and imaginary part of the
amplitude.

From these results, we set up a procedure to test whether it is
possible to use the perturbation technique for a given lagrangian.

\section{Dispersion Relation, Sum Rules and Unitarity}

Because the vector pion form factor $V(s)$ is an analytic function with a cut from
$4m_\pi^2$ to
$\infty$, the $n^{th}$ times subtracted dispersion relation for $V(s)$ reads: 
\begin{equation}
V(s)=a_0+a_1s+...a_{n-1}s^{n-1}+ \frac{s^{n}}{\pi}\int_{4m_\pi^2}^\infty
\frac{ImV(z)dz}{z^{n}(z-s-i\epsilon)}
\label{eq:ff1} 
\end{equation}
where $n\geq 0$ and, for our purpose, the series around the origin is
considered. Because of the real analytic property of $V(s)$, it is real below $4m_\pi^2$. By
taking the real part of this equation, 
$ReV(s)$ is related to the principal part of the 
dispersion integral involving the $ImV(s)$
apart from the subtraction constants $a_n$.

 The
polynomial on the R.H.S. of Eq. (\ref{eq:ff1}) will be referred in the following as the
subtraction constants and the last term on the R.H.S. as
 the dispersion integral (DI). The
evaluation of DI as a funtion of $s$ will be done later. 
  Notice that
$a_n=V^n(0)/n!$ is the coefficient of the Taylor series expansion for
$V(s)$, where
$V^n(0)$ is the nth derivative of
$V(s)$ evaluated at the origin. The condition for  Eq. (\ref{eq:ff1}) to be valid 
was  that, on the real positive s axis, the
limit $s^{-n}V(s)\rightarrow 0$ as $s\rightarrow \infty$. 
The coefficient $a_{n+m}$ of the Taylor's 
series  is given by:
\begin{equation}
a_{n+m} = \frac{1}{\pi}\int_{4m_\pi^2}^\infty
\frac{ImV(z)dz}{z^{(n+m+1)}}\label{eq:an}
\end{equation}
where $m\geq 0$. The meaning of this equation is clear:
 under the above stated assumption,
not only the coefficient $a_n$ can be calculated but all other coefficients $a_{n+m}$
can also be calculated. The larger the value of $m$, the more sensitive is the value of
$a_{n+m}$ to the low energy values of $ImV(s)$. In theoretical work such as in ChPT
approach, to be discussed later, the number of subtraction is such that to make the DI
converges.

The elastic unitarity relation for the pion form factor is $
ImV(s)= V(s)e^{-i\delta(s)}sin\delta(s) $ where $\delta(s)$ is the
elastic P-wave pion pion phase shifts. Below the inelastic
 threshold of $16m_\pi^2$ where
$m_\pi$ is the pion mass,
 $V(s)$ must have the phase of $\delta(s)$ \cite{watson}. It is an
experimental fact that  below $1.3 GeV$ the inelastic effect is very small, hence, to a
good approximation, the phase  of
$V(s)$ is  $\delta$ below this energy scale.

\begin{equation}
ImV(z)  
                        =\mid V(z)\mid\sin\delta(z) \label{eq:ieu} 
\end{equation}
and 
\begin{equation}
ReV(z)  
                        =\mid V(z) \mid\cos\delta(z) \label{eq:reu} 
\end{equation}

where $\delta$ is the strong elastic P-wave $\pi\pi$ phase shifts. Because the real and
imaginary parts are related by  dispersion relation, it is important to know accurately
$ImV(z)$ over a large  energy region. Below 1.3 GeV,
$ImV(z)$ can be determined accurately because the modulus of the vector form factor
\cite{barkov,aleph} and the corresponding P-wave $\pi\pi$ phase shifts are well measured
\cite{proto, hyams, martin} except at very low energy. 

It is possible to estimate the high energy contribution of the dispersion integral by
fitting the asymptotic behavior of the form factor by the expression,
$V(s)=-(0.25/s)ln(-s/s_\rho)$ where $s_\rho$ is the $\rho$ mass squared.

Using Eq. (\ref{eq:ieu}) and Eq. (\ref{eq:reu}), $ImV(z)$ and $ReV(s)$ are determined
directly from experimental data and are shown, respectively, in Fig.1 and Fig.2.

\section{Analysis of the Experimental Data and Test of Dispersion Relation}

In the following, for definiteness, one assumes
 $s^{-1}V(s)\rightarrow 0$ as $s\rightarrow
\infty$ on the cut, i.e. $V(s)$ does not grow as fast as a linear function of $s$. This
assumption is a very mild one because theoretical models assume that the form factor
vanishes at infinite energy as $s^{-1}$. In this case, one can write a once subtracted
dispersion relation for $V(s)$, i.e. one sets $a_0=1$ and $n=1$ in Eq. (\ref{eq:ff1}).

From this assumption on the asymptotic behavior of the form factor,
 the derivatives of the
form factor at $s=0$ are given by Eq. (\ref{eq:an}) with n=1 and m=0.
 In particular one has:
 
\begin{equation}
 <r_V^2> = \frac{6}{\pi}\int_{4m_\pi^2}^\infty
\frac{ImV(z)dz}{z^2}\label{eq:rms}
\end{equation}
where  the standard definition $V(s) = 1 + \frac{1}{6} <r_V^2>s+c s^2 + d s^3+...$ 
is used.
Eq.(\ref{eq:rms}) is a sum rule relating the pion rms radius to the magnitude 
of the time
like pion form factor and the P-wave $\pi\pi$ phase shift measurements.  Using these data,
the derivatives of the form factor are evaluated at the origin:
\begin{equation}
<r_V^2> = 0.45\pm 0.015 fm^2; c = 3.90\pm 0.20 GeV^{-4}; d = 9.70\pm 0.70 GeV^{-6}
\label{eq:rvn}
\end{equation}
where the upper limit of the integration is taken to be $1.7  GeV^2$. By fitting $ImV(s)$
 by the above mentioned asymptotic expression, the
contribution beyond this upper limit is completely negligible.

The only experimental data on the derivatives of the form factor at zero momentum
transfer  is the rms radius of the pion,
$r_V^2= 0.439\pm.008 fm^2$ \cite{na7}. This value is very much in agreement with that 
determined from the sum rules. In fact the sum rule for the
 rms radius gets overwhelmingly
contribution from the $\rho$ resonance as can be seen from Fig.1. The success of the
calculation  of the r.m.s. radius is a first indication that
 causality is respected and also
that the extrapolation procedures to low energy for the P-wave
 $\pi\pi$ phase shifts and for
the modulus of the form factor  are legitimate.

Dispersion relation for the pion form factor is now shown to be 
well verified by the data over a wide energy region. Using $ImV(z)$ as given by Eq.
(\ref{eq:ieu}) together with the once subtracted dispersion relation,
 one can calculate the
real part of the form factor
$ReV(s)$ in the time-like region and also $V(s)$ 
 in the space like region. Because the
space-like behavior of the form factor is not sensitive
 to the calculation schemes, it will
not be considered here. The result of this calculation
 is given in Fig.2. As it can be
seen, dispersion relation results are well satisfied by the data.

\section{Inadequacy of ChPT}

The i-loop ChPT result can be put into the following form, similar to Eq. (\ref{eq:ff1}):
\begin{equation}
V^{pert(i)}(s)= 1 +a_1s+a_2s^2+...+a_is^i+D^{pert(i)}(s) \label{eq:peri}
\end{equation}
where $i+1$ subtraction constants are needed to make the last integral on the RHS of this
equation converges and
\begin{equation}
DI^{pert(i)}(s)=  \frac{s^{1+i}}{\pi}\int_{4m_\pi^2}^\infty
\frac{ImV^{pert(i)}(z)dz}{z^{1+i}(z-s-i\epsilon)} \label{eq:Dperti}
\end{equation}
with $ImV^{pert(i)}(z)$ calculated by the $ith$ loop perturbation scheme.

Similarly to these equations, the corresponding experimental vector form factor
$V^{exp(i)}(s)$ and $DI^{exp(i)}(s)$  can be
 constructed using the same subtraction constants
 as in Eq. (\ref{eq:peri}) but with  the imaginary part replaced by $ImV^{exp(i)}(s)$, 
calculated using Eq. (\ref{eq:ieu}).

The one-loop ChPT calculation requires 2 subtraction constants.
 The first one is given by
the Ward Identity, the second one is proportional to the r.m.s. radius
 of the pion.

 In Fig.
1,  the imaginary part of the one-loop ChPT calculation for the vector
 pion form factor is
compared with the result of the imaginary part obtained from the experimental data. 
It is seen that they differ very much from each other. One  expects therefore that the
corresponding real parts calculated by dispersion relation should be quite different from
each other.

In Fig.2  the full real part of the one loop amplitude is compared 
 with that obtained from experiment. At very low energy
one cannot distinguish the perturbative result from the experimental one due to the
dominance of the subtraction constants. At an energy around $0.56 GeV$ there is a
definite difference between the perturbative result and the experimental data.  This
difference becomes much clearer in Fig. 3 where only the
 real part of the perturbative DI,
$ReDI^{pert(1)}(s)$,  is compared with the corresponding experimental quantity,
 $ReDI^{exp(1)}(s)$.
 It is seen that even at 0.5 GeV the discrepancy is clear.
 Supporters of ChPT would argue
that ChPT would not be expected to work at this energy. One would have to go to a  lower
energy where the data became very inaccurate. 
 
This  argument is false as can be seen by comparing  the ratio
$ReDI^{pert(1)}/ReDI^{exp(1)}$. It is seen in Fig. 4 that \emph{
  everywhere} below 0.6 GeV this ratio
differs from unity by a factor of 6-7 due to the presence of  non perturbative effects.

 Similarly to the one-loop calculation, the  two-loop results are plotted
 in Fig. (1) - Fig.
(4) \cite{Gasser3}.  Although the two-loop result is better
 than the one-loop calculation,
because more parameters are introduced, calculating  higher 
loop effects will not  explain
the data because in ChPT both the form factor and scattering amplitude which enter in
the imaginary part calculation are dominated by a polynomial behavior.

It is seen that perturbation theory is inadequate for the vector pion form factor even at
very low momentum transfer. This fact is due to the very large value of the pion r.m.s.
radius or a very low value of the $\rho$ mass (see below).

\section {Consequences of Unitarized Models}

Two unitarized models which are relevant are as follows.
The first model is obtained by introducing a zero in the calculated  form factor 
 to get an agreement with the experimental r.m.s. radius
. The pion form factor is now multiplied by ${1+\alpha s/s_\rho}$ where 
$s_\rho$ is the $\rho$ mass squared
\cite{Truong4}. 

The experimental data can be  fitted with a $\rho$ mass equal to $0.773 GeV$ and
$\alpha=0.14$. These results are in excellent agreement
with the data \cite{aleph,na7}.

The second model, which  is more complete at the expense of introducing  more parameters,
is  based on the two-loop ChPT calculation with unitarity taken into account. It has the
singularity associated with the two loop graphs. Using the same inverse amplitude method
as was done with the one-loop amplitude, but  generalizing this method to two-loop
calculation, Hannah has recently obtained a remarkable fit to the 
pion form factor in the
time-like and space-like regions. His result is equivalent
 to  the (0,2)  Pad{\'e} approximant
method as applied to the two-loop ChPT calculation \cite{hannah1}.
  Both models contain ghosts which can be shown to be unimportant
\cite{ht}.

As can be seen from Figs. 1, 2 and 3 the imaginary and real
 parts of these two models are
very much in agreement with the data. A small deviation of
 $ImV(s)$  above $0.9 GeV$ is due
to a small deviation of the phases of
$V(s)$ in these two models from the data of the P-wave $\pi\pi$ phase shifts. 

\section{Criteria for the Validity of Perturbation Theory}

Let us examine in details the one-loop ChPT
calculation of the vector pion form factor
$V(s)$ \cite{holstein}:

\begin{equation}
                 V^{pert.}(s) = 1 +\frac{s}{s_{R}}\ + {1\over 96\pi^2f_\pi^2}
((s-4m_\pi^2)
 H_{\pi\pi}({s}) + {2s\over 3} ) \label{eq:pertv}
\end{equation}
where $f_\pi=0.93 GeV$ and the r.m.s. radius of the vector form factor is related to $s_R$
by the definition $V ^{'}(0) =\frac{1}{6} r_V^2 = 1/s_{R}$. The function
$H_{\pi\pi}({s})$ is given by:
\begin{equation}
   H_{\pi\pi}(s) = (2 - 2 \sqrt{s-4m_\pi^2\over s}\ln{{\sqrt{s}+\sqrt{s-4m_\pi^2}\over
2m_\pi}})+i\pi\sqrt{s-4m_\pi^2\over s} \label{eq:H}
\end{equation}
for$ s>4m_\pi^2$; for other values of s, $H_{\pi\pi} (s)$ can be obtained by analytic
 continuation.

The unitarised version of Eq. (\ref{eq:pertv}), obtained by the inverse amplitude, the
Pad{\'e} approximant  or the N/D methods, is given by \cite{Truong1, Truong3}:
 \begin{equation}
         V(s) = \frac{1} {1 -s/s_{R} - {1\over 96\pi^2f_\pi^2}\{(s-4m_\pi^2)
 H_{\pi\pi}({s}) + {2s/3}\}} \label{eq:vu}
\end{equation}

It is obvious that Eq. (\ref{eq:vu}) has the Breit-Wigner resonance character
while that from Eq. (\ref{eq:pertv}) does not, although their amplitude and  first
derivative are identical at $s=0$. Furthermore, if the parameter $s_{R}$ was fixed by the
 the r.m.s. radius, the $\rho$ mass squared, $s_\rho$, would come out to be slightly low
compared measured
$\rho$ mass. Neglecting this last problem which is unimportant here, the Taylor's series expansion around $s=0$ reveals that Eq. (\ref{eq:vu}) gives
rise to a coefficient of the $s^2$  term as 
$(1/s_{R})^2\simeq 4.0 GeV^{-4}$ which is much larger than that coming from the
  third term of Eq. (\ref{eq:pertv}), $1/(960\pi^2m_\pi^2 f_\pi^2) \simeq 0.63
GeV^{-4}$. This is the signal of the failure of the perturbation method.

In other words, ChPT should work if the r.m.s. radius is much smaller than its 
experimental value or $s_R \gg \sqrt{960}\pi f_\pi m_\pi =1.3 GeV^2$. This last
condition means that the physical $\rho$ mass is far too small for the perturbation
theory to be valid. 

We can generalize our criteria for the validity of the perturbation method to the nth
loop result. For this purpose we have to add one more step in our calculation in order
to make it free from criticisms: Eq. (\ref{eq:vu}) could in principle contain unwanted
poles due to the unitarisation procedure. This can be eliminated by subtracting out the
ghost pole which can be conveniently done by writing down  a dispersion relation for
$V(s)$ with the same subtraction constants as used in the perturbative series Eq.
(\ref{eq:pertv}), but with the imaginary part given by Eq. (\ref{eq:vu}) \cite{ht}.

Provided that the ghost removing procedure did not change very much the unitarised
amplitude,  one could then compare the perturbative calculation with the
modified unitarized result. If their difference was negligible, one could  be sure
that the perturbative scheme could then be used.

\section {Further Remarks and Conclusion}

Because the calculable quantities of the vector pion form factor in the ChPT scheme are
too small (well within experimental errors) compared with the uncalculable ones, unless
some unitarisation is made, it may be better to give up this perturbative scheme. One
would then gain in the transparency of the physics.

Although we have not made here a detailed study of the processes $\pi, \eta \rightarrow
\gamma e^+e^-$  the loop contributions can be shown to
be completely negligible compared with the subtracted terms and are therefore
not relevant.

In conclusion, higher loop perturbative calculations
 do not solve the unitarity problem. The perturbative
scheme has to be supplemented by the well-known unitarisation 
schemes such as the inverse
amplitude, N/D and  Pad{\'e} approximant methods
 \cite{Truong1, Truong3,  hannah1,  ht,
Truong5, lehmann}.

\newpage

{\bf Figure Captions}

Fig.1~: The imaginary part of the vector pion form factor $ImV$, given by Eq.
(\ref{eq:ieu}), as a function of  energy in the $GeV$ unit.
 The solid curve  is the the
experimental results with experimental errors; the long-dashed 
curve is the two-loop ChPT
calculation, the medium long-dashed curve is the one-loop
 ChPT calculation, the short-dashed
curve is from the modified unitarized one-loop ChPT calculation 
fitted to the $\rho$ mass
and the experimental r.m.s. radius, and the dotted curve is the unitarized two-loop
calculation of Hannah
\cite{hannah1}.

Fig. 2~: The real parts of the pion form factor $ReV$, given by Eq. (\ref{eq:reu}) as a
function of energy. The curves are as in Fig. 1. The real part of the
form factor calculated by the once subtracted dispersion relation
using the experimental imaginary part is also given by the solid line.

Fig. 3~: The real parts of the dispersion integral ReDI as a function of energy.
 The curves are as in Fig. 1.

Fig. 4~:The ratio of the one-loop ChPT to the corresponding experimental quantity,
$ReDI^{pert(1)}/ReDI^{exp(1)}$, defined by Eq. (\ref{eq:Dperti}), 
as a function of energy, 
is given by the solid line; the corresponding ratio for 
the two-loop result is given by the
dashed line. The ratio of the unitarized models to 
the experimental results is unity (not
shown). The experimental errors are estimated to be less than 10\%.
\newpage
\begin{figure} 
\epsfbox{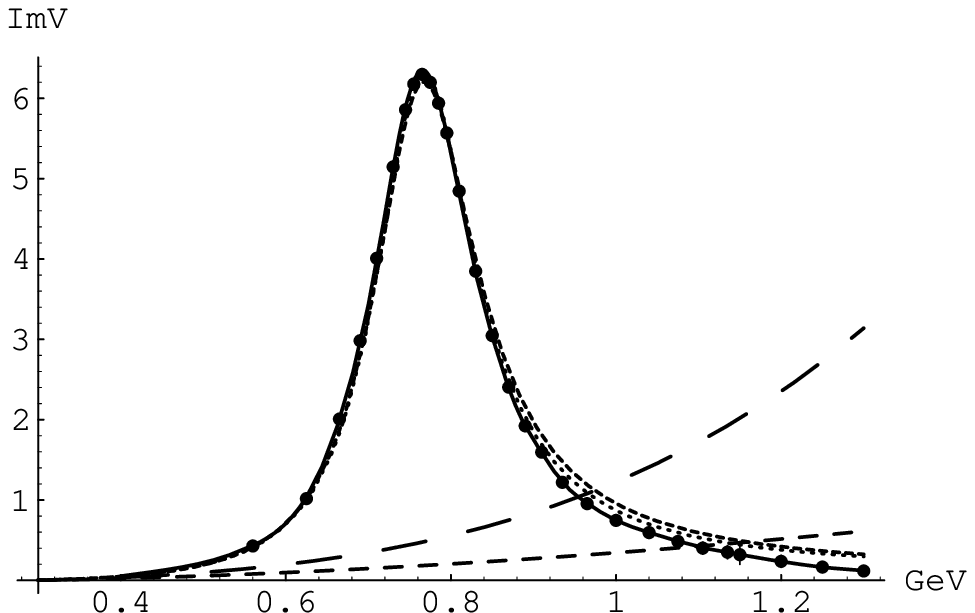}
\caption{}
\label{Fig.1}
\end{figure}

\begin{figure}
\epsfbox{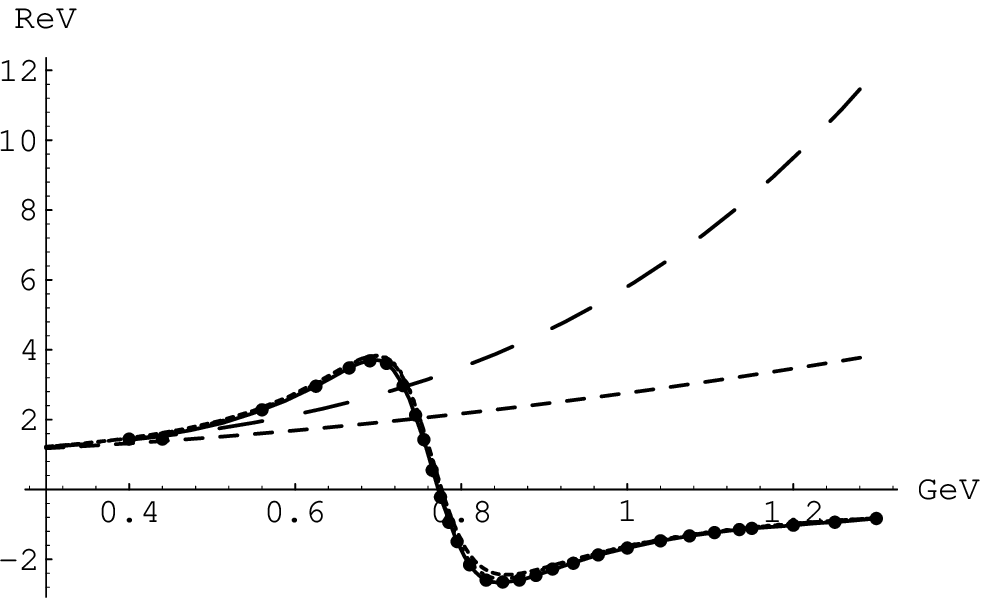}
\caption{}
\label{Fig.2}
\end{figure}
\newpage
\begin{figure}
\epsfbox{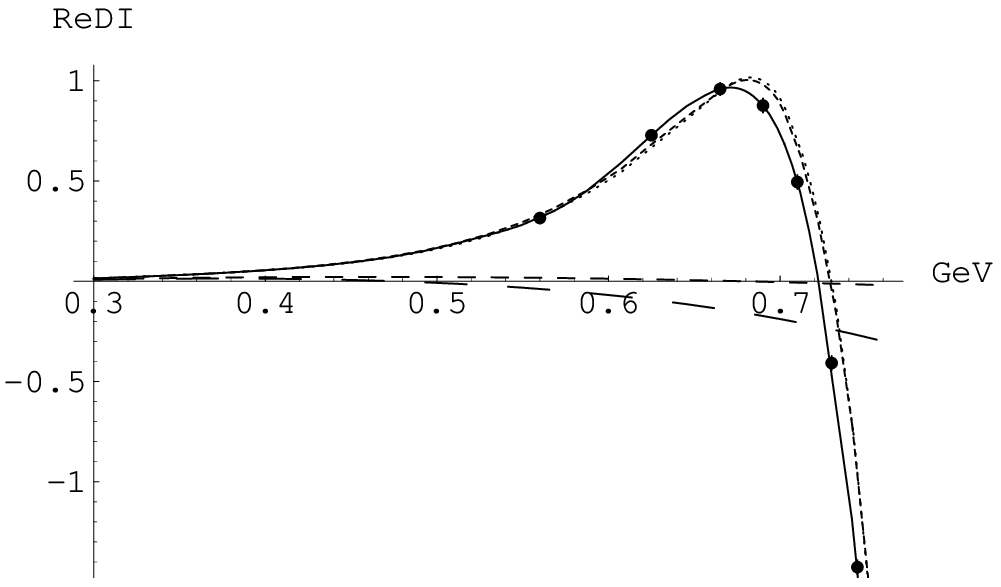}
\caption{}
\label{Fig.3}
\end{figure}

\begin{figure}
\epsfbox{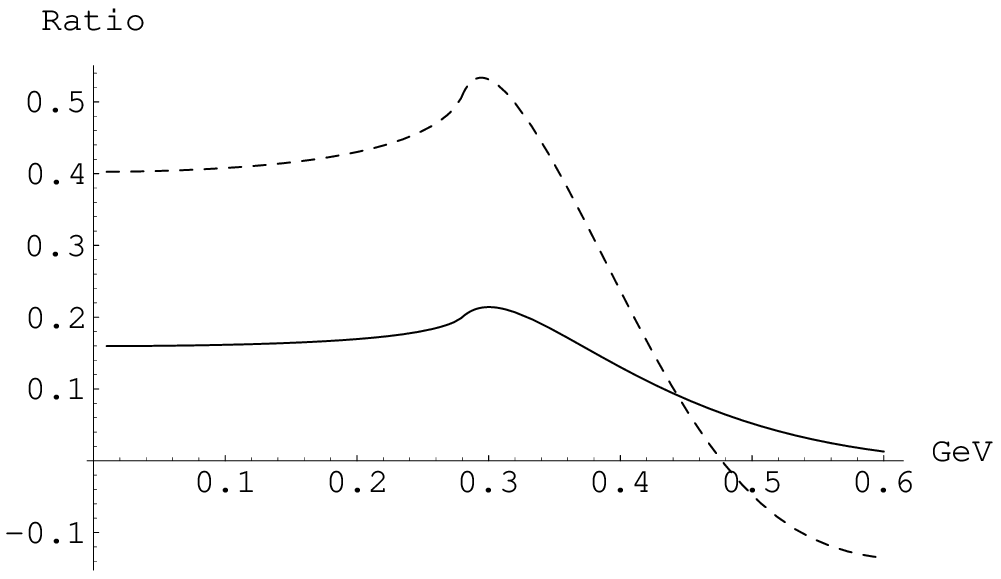}
\caption{}
\label{Fig.4}
\end{figure}

\end{document}